\documentclass[twoside]{article}
\usepackage{fleqn,espcrc2,epsfig}

\usepackage{graphicx}
\usepackage[figuresright]{rotating}

\newcommand{\AmS}{{\protect\the\textfont2
  A\kern-.1667em\lower.5ex\hbox{M}\kern-.125emS}}

\title{$F_B$ from moving $B$ mesons}

\author{S.Collins\address[GLA]{Department of Physics and Astronomy,
        University of Glasgow, Glasgow, G12 8QQ, UK}, 
        G.Cowan\addressmark,
        C. T. H. Davies\addressmark,
        J. Hein\address{Department of Physics and Astronomy,
         University of Edinburgh, Edinburgh, EH9 3JZ, UK},
        G. P. Lepage\address{Newman Laboratory of Nuclear Science, 
         Cornell University, Ithaca NY 14853, USA},
        C. Morningstar\address{Department of Physics, Carnegie Mellon University, 
        Pittsburgh PA 15213, USA}
        and
        J. Shigemitsu\address[OSU]{Department of Physics, the Ohio State 
                University, OH 43210 USA}.
        }

\newcommand{\figwidth}{6.85cm}
\begin{document}

\begin{abstract}
We show results for the $B$ meson decay constant calculated both for $B$ mesons 
at rest and those with non-zero momentum and using both the temporal and 
spatial components of the axial vector current. It is an important check of 
lattice systematic errors that all these determinations of $f_B$ should agree. 
We also describe how well different smearings for the $B$ meson work 
at non-zero momentum - the optimal smearing has a narrow smearing on the $b$ quark.  
\vspace{1pc}
\end{abstract}

\maketitle

\section{Introduction}

Matrix elements involving moving $B$ and $D$ mesons are important for 
studies of $B \rightarrow D$ and $B \rightarrow \pi$ decay.  It is 
necessary to understand the systematic errors in lattice QCD that come from 
the presence of non-zero momenta and the optimal way in which 
to handle such mesons on the lattice. An easy place to study these 
effects is in the determination of the $B$ meson decay constant, $f_B$~\cite{simone}. 

In the absence of discretisation errors it should be true that 
\begin{equation}
<0|A_{\mu}|B> = f_B p_{\mu}.
\end{equation}
We use both $A_0$ and $A_k$ with all current corrections and 
renormalisation through $\alpha_s/M_b$~\cite{junko_colin}. 

Results shown are for an ensemble of 278 $12^3\times24$ 
lattices at $\beta$ = 5.7. Clover light quarks are used with $\kappa =
\kappa_s$ and NRQCD heavy quarks are used with masses in 
lattice units of 2 and 8 ($m_ba$ = 4.0). All momenta are 
considered up to $(pa)^2$ of 16 in units of $(2\pi/L)^2$,
i.e. roughly 4 ${\rm GeV}^2$ at this lattice spacing. 

\section{Smearing}
We generated heavy and light quark propagators with 3 
different smearings: a delta function, a narrow 
Gaussian (width 1) and a broad Gaussian (width 3). 
We combined these smearings together to make 6 different 
smearings for the heavy-light meson and analysed 
the resulting smeared-smeared meson correlators using 
the constrained curve fitting methods described in \cite{gpl}.
The results are shown in Figure 1, which plots the 
amplitude of the ground state $B$ meson as a function 
of squared momentum for some of these smearings. 
It is clear that the optimal smearing for a moving 
heavy-light meson is one in which the smearing on the heavy 
quark is a narrow one and the smearing on the light 
quark is a broad one. This is not surprising if 
one considers that 
the heavy quark is carrying almost all 
of the meson momentum. The overlap as a function of momentum 
is maximized if the heavy quark has a delta function smearing,
but in that case the statistical noise is large because 
of the well-known problem that an unsmeared heavy quark
has a poor signal/noise ratio~\cite{oldgpl}. 
At zero momentum this has led to the received wisdom that the 
heavy quark should have a broad smearing - Figure 1 shows that 
this is not correct at non-zero momentum. 

\begin{figure}
\centerline{\epsfig{file=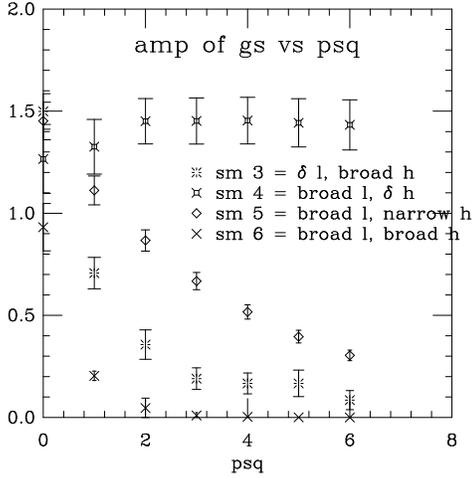,width=\figwidth}}
\caption[kgkk]{The amplitude for the ground state in 
various heavy-light smeared-smeared correlators as a function of $\vec{p}^2$ in 
units of $(2\pi/L)^2$. The heavy mass was $ma$ = 8.}
\end{figure}

Figure 2 amplifies this point by showing the  rapid plateau 
and consequent small error on the kinetic energy 
possible with a good finite-momentum
smearing, in contrast to that obtained with a smearing that 
might have been considered a good one at zero momentum. 

\begin{figure}
\centerline{\epsfig{file=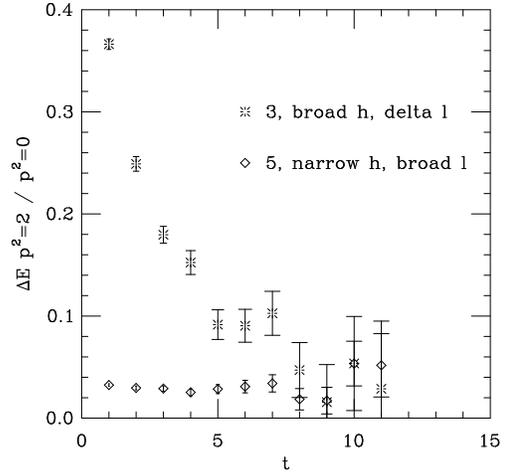,width=\figwidth}}
\caption[peigp]{The effective energy splitting between 
smeared-smeared correlators of $\vec{p}^2$ = 0 and 2 in units 
of $(2\pi/L)^2$ against time, for smearings 3 and 5 of Figure 1. }
\end{figure}

\section{$f_B$ from $A_0$}

We used smearing `5' from the above analysis and the 
constrained curve fitting methods to determine
$f_B$ for moving $B$ mesons and the temporal 
axial current. Matrix elements for $A_0$ are 
obtained from a simultaneous fit to {\it all}
$A^i_0$ (i=0,1,2) and $A^i_k$ (i=0,1,2,3,4) current correlators at a given 
momentum. The construction of the continuum $A_0$ from 
$A^{i,latt}_0$ is described in \cite{junko_colin}.

\begin{figure}
\centerline{\epsfig{file=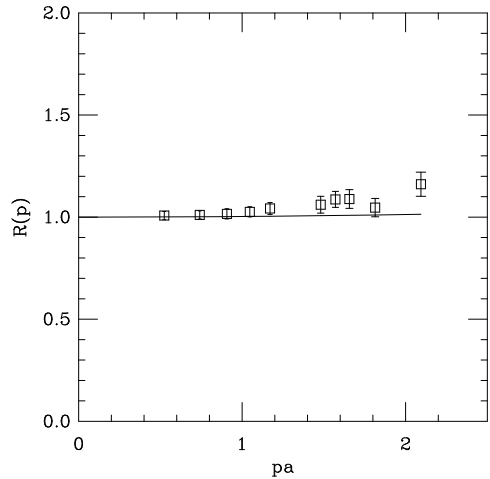,width=\figwidth}}
\caption[wrwd]{The ratio $R(p)$ (see text) for $A_0$ 
as a function of momentum, $pa$.
The curve is $\sqrt{E/M}$ where $E$ is the heavy-light meson energy 
and $M$ its mass. }
\end{figure}

Figure 3 shows results for the ratio:
\begin{equation}
R(p) = \frac{\langle 0| A_0 | B(p) \rangle / \sqrt{E}} {\langle 0| A_0 | B(0) \rangle  /\sqrt{M}}
\end{equation}
for $ma$=8 using $\alpha_s(2/a)$ in the renormalisation. 
The expected curve $\sqrt{E/M}$ is also shown. 
No disagreement is seen until 
the highest momentum.  At $ma$ = 2, larger discrepancies 
appear~\cite{inprep}. 

\section{$f_B$ from $A_k$}

The matrix elements for the spatial axial current 
behave rather differently to those for the temporal
axial current since they must vanish when there is 
no component of $\vec{p}$ along the direction of the 
current. Even the leading order current, $A^0_k$ = 
$\overline{q}\gamma_5 \gamma_k Q$ then has a matrix 
element $\cal{O}$$(p_k/M)$ with respect to $A^0_0$.  
This is of the same order as the `$1/M$' suppressed 
current contributions from $A^1_k$ = $\overline{q}
\gamma_5 \gamma_k (\vec{\gamma}\cdot \vec{D}/2M) Q$
and $A^3_k$ = $\overline{q} \gamma_5 (\vec{D}/2M) Q$.
$A^1_k$ contributes at tree level and $A^3_k$ at
one-loop to the final result for $A_k$. To understand 
the size of different current contributions it is important 
to use a power-counting in both the external velocity
and in $\Lambda_{QCD}/M$~\cite{inprep}.

\begin{figure}
\centerline{\epsfig{file=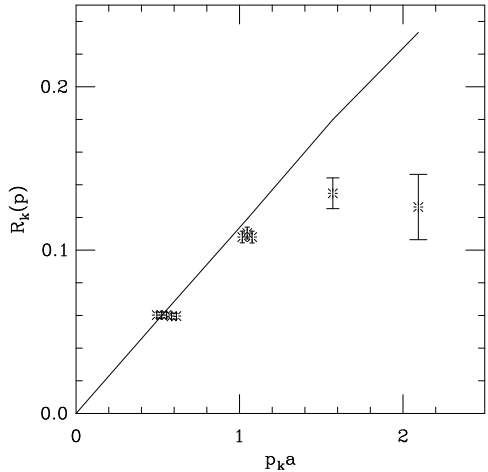,width=\figwidth}}
\caption[wrwd]{The ratio $R_k(p)$ (see text) for $A_k$ 
as a function of momentum, $p_ka$. }
\end{figure}

Figure 4 shows results for the ratio:
\begin{equation}
R_k(p) = \frac{\langle 0| A_k | B(p) \rangle } {\langle 0| A_0 | B(p) \rangle }
\end{equation}
as a function of  $p_k$ for a range of values of 
$\vec{p}^2$, for $ma$=8.  The line shown is $p_k/E$ (variation 
with $\vec{p}^2$ of this line is not significant). Good agreement 
with the line is found for $p_ka$ = 1,2 ($\times 2\pi/L$) at 
all $\vec{p}^2$ but $p_ka$ = 3,4 show signs of deviation, 
presumably from missing current corrections that are higher order in the 
external velocity. 

Figure 4 shows agreement between $f_B$ from $A_0$ and $A_k$ 
which resolves a long-standing problem. Ref. \cite{simone}
found a disagreement of $\cal{O}$(10\%) when comparing the 
tree-level matrix elements for clover fermions. 
This ignores the contribution of $A^3_k$ and does not allow
for the clover equivalent of the different renormalisation of $A^0_{0,k}$ and 
$A^1_{0,k}$. Both these effects are included here. 

\section{Conclusions}
We have demonstrated the usefulness of optimal 
(i.e. narrow) smearing for heavy quarks in moving 
heavy-light mesons, and of constrained curve fitting 
in extracting results out to much higher momenta 
than have previously been attempted. 
Good agreement is found for $f_B$ 
at zero and non-zero momentum. For the temporal 
axial current at large $ma$ this holds out to the 
highest momenta studied. To understand the 
behaviour of the currents at non-zero momentum 
it is important to use a power-counting appropriate 
to moving heavy-light mesons. 

\vspace{2mm}

{\bf Acknowledgements}
SC, GC, CD and JH are members of the UKQCD collaboration. 
We are grateful to the following bodies for support of 
this work: PPARC, DoE, NSF and the EU under 
HPRN-2000-00145 Hadrons/Lattice QCD.

\end{document}